# On-chip three-dimensional high-Q microcavities fabricated by femtosecond laser direct writing

Jintian Lin[1,3], Shangjie Yu[2], Yaoguang Ma[2], Wei Fang[2,*], Fei He[1], Lingling Qiao[1], Limin Tong[2], Ya Cheng[1,†], Zhizhan Xu[1,‡]

[1] *State Key Laboratory of High Field Laser Physics, Shanghai Institute of Optics and Fine Mechanics, Chinese Academy of Sciences, P.O. Box 800-211, Shanghai 201800, China*

[2] *State Key Laboratory of Modern Optical Instrumentation, Department of Optical Engineering, Zhejiang University, Hangzhou 310027, China*

[3] *Graduate School of Chinese Academics of Science, Beijing 100039, China*

[*]*Email: wfang08@zju.edu.cn*

[†]*Email: ya.cheng@siom.ac.cn*

[‡]*Email: zzxu@mail.shcnc.ac.cn*

**Abstract:**

We report on the fabrication of three-dimensional (3D) high-Q whispering gallery microcavities on a fused silica chip by femtosecond laser microfabriction, enabled by the 3D nature of femtosecond laser direct writing. The processing mainly consists of formation of freestanding microdisks by femtosecond laser direct writing and subsequent wet chemical etching. $CO_2$ laser annealing is followed to smooth the microcavity surface. Microcavities with arbitrary tilting angle, lateral and vertical positioning are demonstrated, and the quality (Q)-factor of a typical microcavity is measured to be up to $1.07\times10^6$, which is currently limited by the low spatial resolution of motion stage used during the laser patterning and can be improved with motion stages of higher resolutions.

## 1. Introduction

Whispering gallery mode (WGM) optical microcavities, that trap light via the total internal reflection at the circular boundary formed between the dielectric cavity and surroundings, exhibit very high quality (Q)-factors and very small volume. These excellent properties lend them very attractive for a variety of fields, ranging from fundamental science to engineering applications [1]. Due to the nature of WGM, the light emission from the cavity is mainly in the plane parallel to the optical mode, i. e., for the on-chip WGM microcavities such as microdisks [2], microtoroids [3], and deformed microcavities [4] fabricated by lithography method, parallel to the substrate. However, this makes an obstacle for many applications that prefer out-of-plane light coupling or extraction. Though Levi et. al. had already started an early approach [5] shortly after their demonstration of the first microdisk lasers, till now, the realization of out-of-plane light output from a microdisk laser is still limited to vertical emission by adding grating structures either at the boundaries [5] or on the top surfaces [6] of microdisks.

Recently, femtosecond laser micromachining has been proved as a promising solution for high-precision and flexible fabrication of three dimensional (3D) microstructures, such as microoptics [7,8], microfluidic laser [9], hollow waveguide [10], microfluidic channels [11] and polymer-based microcavities [12-14]. In comparison with polymer, fused silica is considered to be a very attractive substrate due to its wide transparency range and extremely low intrinsic material absorption loss [15]. 3D microoptics in fused silica can be fabricated by femtosecond laser direct writing, followed by wet chemical etching and a postannealing method [10].

In this article, by applying femtosecond laser direct writing method, we demonstrate a new way to realization of 3D high-Q microcavities on fused silica wafer that may have the light output from optical mode on arbitrary planes respect to substrate plane. Either the tilting angles or the heights of microcavities are free of limitations, which, to the best of our knowledge, can not be realized with any planar lithographic fabrication techniques. After the annealing process, the Q-factor of a microcavity is measured to be above $10^6$, which can be further improved by replacing the low-resolution (~1 μm) motion stage used in the process of femtosecond laser fabrication with a better one, as explained in details in main context.

**2. Fabrication of 3D microtoroidal cavities with femtosecond laser direct writing**

In this work, commercially available fused silica glass substrates (UV grade fused silica JGS1 whose upper and bottom surfaces are polished to optical grade) with a thickness of 1 mm are used. The process flow for fabrication of the ultra-high-Q microcavity mainly consists of two steps: (1) femtosecond laser exposure followed by selective wet etching of the irradiated areas to create the microdisk structures; and (2) selective reflow of the silica cavities by $CO_2$ laser annealing to improve the quality factors, as illustrated in Fig. 1. The laser system consists of a Ti: sapphire oscillator (Coherent, Inc.) and a regenerative amplifier, which emits 800 nm, ~58 fs pulses with maximum pulse energy of ~5 μJ at 250-kHz repetition rate. The initial 8.8-mm-diameter beam is reduced to 5 mm by passing through a circular aperture to guarantee a high beam quality. Power adjustment is realized using neutral density (ND) filters. The glass samples can be arbitrarily translated in 3D space with a resolution of 1 μm by a PC-controlled XYZ stage. In the femtosecond laser direct writing, a 100× objective with a numerical aperture (NA) of 0.9 is used to focus the beam down to a ~1 μm-dia. spot, and the average femtosecond laser power measured before the objective is ~0.05 mW. To form the microdisk supported by a thin pillar, a layer-by-layer annular scanning method with the lateral scanning step set to be 1 μm is adopted, and the femtosecond laser scanning speed is chosen to be ~600 μm/s. The scanning is designed to modify the regions surrounding the areas which forms a disk with a radius of 29 μm, a thickness of ~7 μm, tilted with respect to the substrate at 24°, and an underneath pillar with a radius of 12 μm, as the modified regions will be preferentially etched away in hydrofluoride (HF) acid. In addition, the angle between the disk and the pillar is set to be 57°.

After the laser exposure, the sample is subjected to a ~20 min bath in a solution of 5% HF diluted with water, until the laser-modified part is etched away, leaving an on-chip microdisk structure [Fig. 2(a)]. To facilitate the further coupling of light into the microtoroid with a fiber taper, the microdisk is located close to one corner of substrate, and both the adjacent sidewalls next to the microdisk are polished to have the edge of the micodisk exposed. To improve the surface smoothness and achieve the desirable high cavity Q factor, we smooth the surface by introducing surface-normal-irradiation with a $CO_2$ laser (Synrad Firestar V30). The $CO_2$ laser beam is focused by a lens to a circular spot approximately 100 μm in diameter. Because of the strong absorption of fused silica in the infrared band, the microdisk can be heated up by the laser to melting temperature which causes reflow. The laser is operated with a repetition rate of 5 kHz, and the time-averaged beam intensity is controlled by adjusting the duty ratio. As the disk diameter thermally shrinks during the reflow, surface tension induces a collapse of the fused silica disk, leading to a toroidal-shaped boundary [3]. During this process the disk is monitored by a charge coupled device (CCD) from the side with a 60× objective lens. The total reflow process takes merely ~4 s, with a duty ratio of 5.0%. Due to the surface tension, the smoothness of the microtoroid surface is excellent, as shown in optical micrograph (Fig. 2(b)) and scanning electron microscope (SEM) image (Fig.2(c)). The overall disk diameter is reduced to 43 μm with a 9-μm-thick toroid-shaped boundary. The freedom in controlling the lateral and vertical positions of the microcavities is demonstrated in Fig. 2(d), where two microtoroids are located close by yet with different heights. The heating $CO_2$ laser beam is focused to a relatively small spot so that the reflow process on one of the microcavity will not affect the other. Thus the shape of each microtoroid can be controlled by adjusting the exposure time individually. The left microtoroid in Fig. 2(d) is achieved with 4-second longer $CO_2$ laser annealing time in the reflow process than the right one in Fig. 2(d).

## 3. Characterization of the microcavity and discussion

To characterize the mode structure and Q factor of the microtoroidal cavity, resonance spectra are measured via the optical fiber taper coupling method [16]. For facilitating a convenient coupling, a microtoroidal cavity parallel to the substrate with a diameter of ~39 μm and a thickness of ~9 μm is fabricated using the technique mentioned in Sec. 2, as shown by its SEM image in Fig. 3(a). A swept-wavelength tunable external-cavity diode Laser (New Focus, Model: 6528-LN) and a swept spectrometer (dBm Optics, Model: 4650) are used to measure the transmission spectrum from the fiber taper with a resolution of 0.1pm. A periodic pulse signal with a power of 1 dBm is used to continuously sweep from 1530 to 1565 nm. The fiber taper formed by heating and stretching a section of a commercial optical fiber (Corning, SMF-28) has a minimum waist diameter of approximately ~1 μm, providing an evanescent excitation of WGMs of the cavity. The microcavity sample is fixed on a three-axis nanopositioning stage with a spatial resolution of 50-nm in the XYZ directions, so that the critical coupling may be realized by carefully adjusting the relative position between the cavity and the fiber taper. We use dual CCD cameras to simultaneously image microcavity and fiber taper from the side and the top, as shown in Fig. 3(b) [side view image not shown]. Please note that the thinnest portion of the fiber taper (dia. ~1 μm) is not very clear in Fig. 3(b) because of the limited resolution of the optical microscope (a 50X objective with a NA of 0.3 is used for the imaging in this case to obtain a sufficiently long working distance of 38 mm).

A resonance transmission spectrum of a fiber taper coupled to the microtoroidal cavity with various excited WGMs is depicted in Fig. 3(c). The experimentally measured free spectral range (FSR=13.65 nm, defined as the wavelength spacing between modes with successive angular mode number) agrees well with the numerical calculation based on the experimentally measured cavity diameter, which is given approximately by the well-known expression [13]:

$$\Delta\lambda_{FSR} \approx \lambda_0^2 / 2\pi R n \tag{1}$$

Where $\lambda_0$ is the wavelength in vacuum, $R$ the radius of the microcavity, $n$ the refractive index of the fused silica. For $R$=19.5μm, $n$=1.445, the theoretically predicted value of FSR is about 13.30 nm at 1534.72 nm. Figure 3(d) shows an individual WGM located at 1534.72 nm with a Lorentzian shaped dip. The linewidth is measured as 1.44 pm, and the Q factor for the mode is calculated to be $1.07\times10^6$, inferred from the Lorentzian fit of the spectrum, as shown in Fig. 3(d). This indicates that femtosecond laser micromachining on fused silica enables fabrication of smooth cavity surfaces with low surface-scattering loss of the WGMs..

It is noteworthy that microtoroids reported in previous publications can frequently achieve ultra-high-Q in the order of over $10^8$ [3], which is two orders of magnitude higher than the Q achieved in the current experiment. We believe that the major reason for this relatively lower-Q measured in our case is the imperfection in the fabrication of the microdisk with our motion stage of limited resolution (~1 μm) prior to the annealing by the $CO_2$ laser. As we have seen in Fig. 2(a), the edge of the microdisk shows some irregular features prior to the annealing by the $CO_2$ laser. After the reflow process, though the small scale roughness has been removed, as we can see very smooth surface for all the samples under SEM with high magnifications, on some of the microcavities we do have observed clear large scale deformation so that boundary of the cavitie deviates from an ideal circle. Such deformation is introduced by the non-uniformity of the disk radius or/and thickness after wet chemical etching. This problem can be solved by replacing the current motion stage with a higher resolution one. Actually for many applications (i. e., on-chip microcavity lasers or some sensors), the currently achieved Q value in the order of $10^6$ will already be sufficient.

Fabrication of a microcavity by femtosecond laser direct writing surely requires more time than traditional photolithography; however, it should be stressed here that it is very difficult, if not completely impossible, to realize such flexible 3D microcavities presented here for planar lithography method such as photolithography or electron beam lithography. Though the microtoroids still exhibit isotropic properties of the confined WGM, the femtosecond laser direct writing can easily create preferred deformation into the cavity and thus introduce directional light output into any desired solid angle. This would open a broad spectrum of applications such as micro-lasers, sensors, and so on.

## 4. Conclusion

To summarize, we demonstrate the fabrication of 3D microcavities on the fused silica chip by femtosecond laser direct writing. The typical Q factors of the microcavities are over $10^6$. Further improvement of the Q-factor should be achievable by replacing the XYZ stage of relatively low resolution in the current writing system with a high-resolution XYZ stage. Though the cavity shown here is a passive device, with such high Q factors, lasing can be easily achieved by coating the cavity surface with gain medium such as semiconductor colloidal quantum dots, or by using rare-earth-doped substrate to provide optical gain. Therefore, our technique opens a new avenue for constructing high-Q microcavities with non-inplane geometries which may find important uses in both fundamental research and biological and chemical sensing applications.

**Acknowledgement**

We thank Prof. Yunfeng Xiao of Peking University, Dr. Xiaoshun Jiang of Nanjing University for the helpful discussion. The work is supported by National Basic Research Program of China (No. 2011CB808100), NSFC (Nos. 11134010, 60825406, 61008011, 10974213, 11104245) and Fundamental Research Funds for the Central Universities.

**Figure captions:**

Fig. 1 (Color online) Procedures of fabrication of 3D microcavity by femtosecond laser direct writing.

Fig. 2 (Color online) Optical microscope images of a tilted fused silica microdisk fabricated by femtosecond laser micromachining and HF wet etching (a) before and (b) after $CO_2$ laser annealing. Insets in (a) and (b): top view of the microcavity, (c) SEM images of the cavity, and (d) SEM image of two microtoroidal cavities with different heights after $CO_2$ laser annealing.

Fig. 3 (Color online) (a) SEM image of a microtoroidal cavity parallel to the substrate whose Q factor is to be examined. (b) An optical micrograph of the microtoroidal cavity coupled with a fiber taper. (c) Transmission spectrum of the microcavity coupled with the fiber taper. The free spectral range of 13.65nm agrees well with the numerical calculation result. (d) Lorentzian fit (red solid line) of measured spectrum around the resonant wavelength at 1534.72nm (black dotted line), showing a Q factor of $1.07 \times 10^6$.

Fig. 1

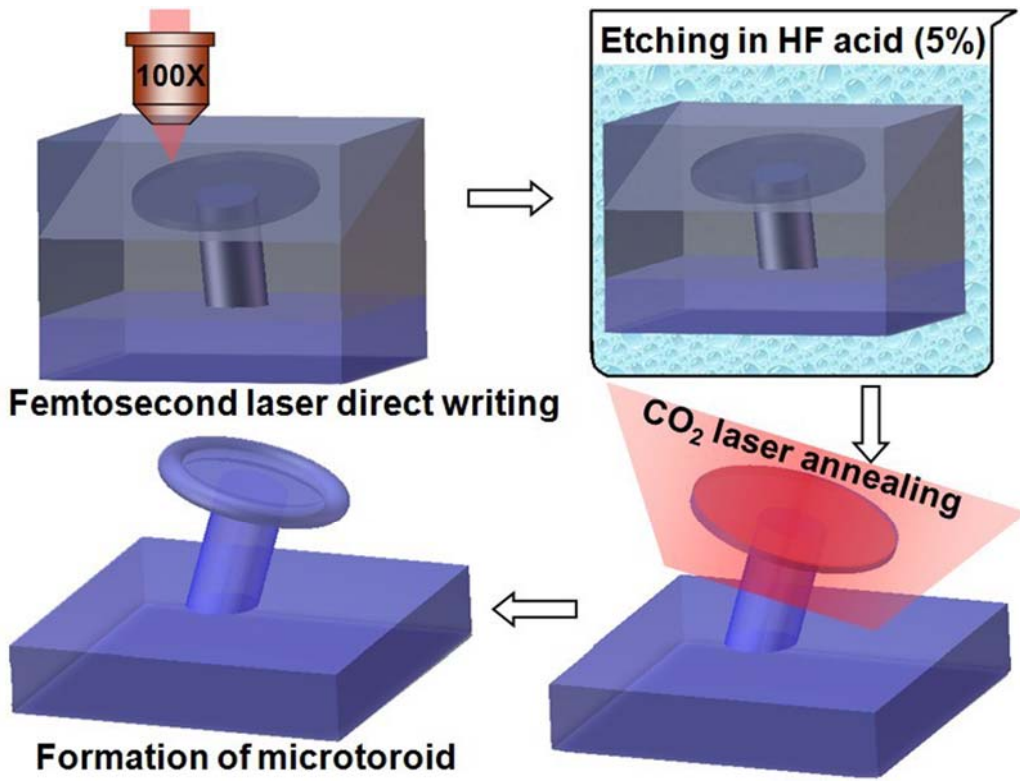

Fig. 2

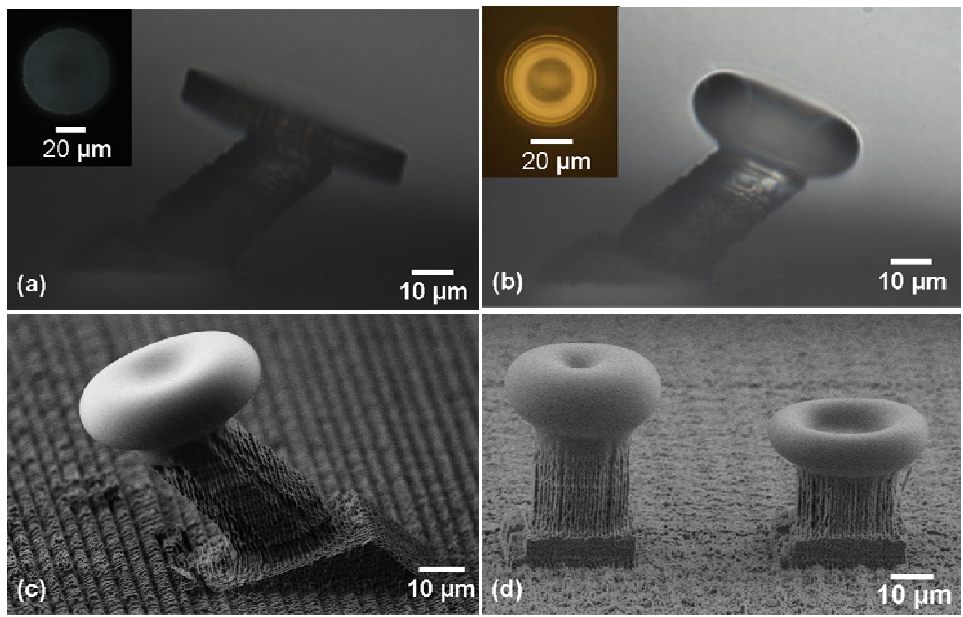

Fig. 3

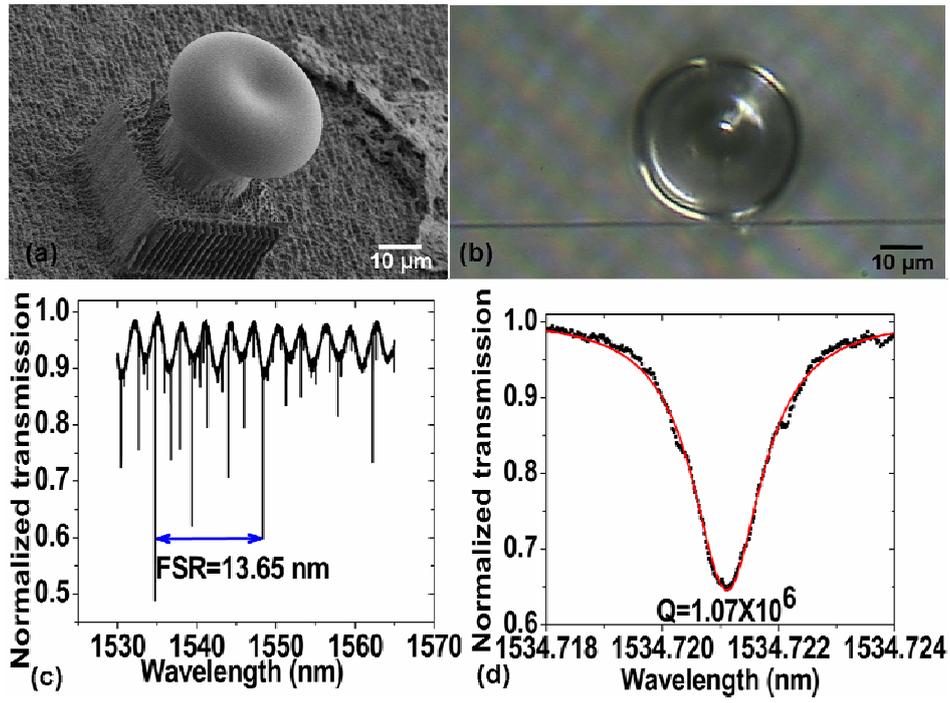